\documentclass[12pt,amsmath,amssymb]{article}

\usepackage{graphicx}
\usepackage{amsmath}
\usepackage{amssymb}
\usepackage{epsfig}
\usepackage{wasysym}
\usepackage{bbm}

\usepackage{times}
\newcommand\journal[4]{#1 {\bf #2}, #4 (#3).}
\newcommand\vect[1]{\mathbf{#1}}

\title{The Electron Pairing  of K$_x$Fe$_{2-y}$Se$_2$}
\author{Fa Wang$^{1}$, Fan Yang$^{2}$, Miao Gao$^{3}$, Zhong-Yi Lu$^{3}$,Tao Xiang$^{4,5}$,  and  Dung-Hai Lee$^{6,7}$\\
\normalsize{1. Department of Physics, Massachusetts Institute of Technology, Cambridge, MA 02139, USA,}\\
\normalsize{2. Department of Physics, Beijing Institute of Technology, Beijing 100081, P.R.China,}\\
\normalsize{3. Department of Physics, Renmin University of China, Beijing 100872, China,}\\
\normalsize{4. Institute of Theoretical Physics, Chinese Academy of Sciences, Beijing 100190, China,}\\
\normalsize{5. Institute of Physics, Chinese Academy of Sciences, Beijing 100190, China,}\\
\normalsize{6. Department of Physics, University of California at Berkeley, Berkeley, CA 94720, USA,}\\
\normalsize{7. Materials Sciences Division, Lawrence Berkeley National Laboratory, Berkeley, CA 94720, USA}\\}

\begin{document}

\baselineskip24pt

\maketitle

\begin{abstract}

We studied the pairing instabilities in K$_x$Fe$_{2-y}$Se$_2$ using a two stage functional renormalization group (FRG) method.
Our results suggest the leading and subleading pairing symmetries are nodeless $d_{x^2-y^2}$ and nodal extended $s$ respectively.
In addition, despite having no Fermi surfaces we find the buried hole bands make important contributions to the final effective interaction.
From the bandstructure, spin susceptibility and the FRG results
we conclude that the low energy effective interaction in K$_x$Fe$_{2-y}$Se$_2$ is well described by
a $J_1-J_2$ model with dominant nearest-neighbor antiferromagnetic interaction $J_1$
(at least as far as the superconducting pairing is concerned).
In the end we briefly mention several obvious experiments to test whether the pairing symmetry is indeed $d_{x^2-y^2}$.

\end{abstract}

\section{Introduction}
Very recently a new wave of excitements occurred in the field of iron-based superconductors.
This is stirred up by the discovery of K$_x$Fe$_{2-y}$Se$_2$\cite{Kfese}.
These compounds are isostructural to the ``122''-family of iron pnictides, {e.g.} BaFe$_2$As$_2$,
with the highest transition temperature $T_c\approx$ 33K among iron chalcogenides.
The reason these compounds attracted considerable interests is not their ``high $T_c$''.
Rather it is because they have a very different electronic structure from all other iron-based superconductors.
In particular, the hole pockets near the Brillouin zone center in all other iron-based superconductors
are found to be absent in the highest $T_c$ K$_x$Fe$_{2-y}$Se$_2$\cite{dl,ding}.
The result of ref.~\cite{dl} suggests a very uniform ($\approx 10$meV) superconducting gap around the electron pockets.

The absence of hole pockets is expected from valence count: KFe$_2$Se$_2$ should have $0.5$ doped electron per Fe relative to the parent iron pnictides.
In view of the wide spread belief that the scattering between the hole and electron pockets is crucial for the high pairing scale in the iron-based superconductors\cite{mazin,kuroki,rg1,seo,rg2,rg3,rg4,rg5,rg6,rg7,rg8,flex1,flex2,rpa1,rpa2},
it is surprising that a material without hole pocket can support such a high $T_c$.
It is fair to say the relatively high pairing scale in the absence of hole pockets calls for a re-evaluation of the spin fluctuation pairing mechanism.
This is so because in the other extreme KFe$_2$As$_2$, which has $0.5$ doped holes per Fe and only hole pockets, is a very low $T_c$ ($\approx 3K$)\cite{kfeas1,kfeas2} superconductor with experimental evidences of gap nodes \cite{nodes}.

In this paper we apply the functional renormalization group (FRG) method\cite{rg1,rg3,rg8} to study the pairing instability of K$_x$Fe$_2$Se$_2$.
When the electron-electron interaction is weak compared with the bandwidth, this method is unbiased.
It sums all virtual one-loop scattering processes including particle-hole, particle-particle and vertex corrections.
As other iron-based superconductors, the strength of electron-electron interaction in  K$_x$Fe$_{2-y}$Se$_2$ is uncertain.
Ideally, we should combine FRG with the variational Monte-Carlo calculation\cite{fan}, which is underway.

\section{Model}

According to an earlier DFT result\cite{shein}, KFe$_2$Se$_2$ has cylindrical electron pockets around $(\pi,0,k_z)$ and $(0,\pi,k_z)$ as well as a 3D electron pocket centered around $(0,0,\pi)$.
In our study we uses a two dimensional(2D) five-band tight binding model to describe the $k_z=0$ plane of the bandstructure.
It is obtained via the maximally localized Wannier function fit\cite{wannier} to our own DFT calculation.
In our DFT calculation the plane wave basis method was used \cite{pwscf}.
We adopted the generalized gradient approximation (GGA) of Perdew-Burke-Ernzerhof formula \cite{pbe} for the exchange-correlation potentials.
The ultrasoft pseudopotentials \cite{vanderbilt} were used to model the electron-ion interactions.
After the full convergence test, the kinetic energy cut-off and the charge density cut-off of the plane wave basis were chosen to be 800eV and 6400 eV, respectively.
The Gaussian broadening technique was used and a mesh of $18\times 18\times 9$ k-points were sampled for the Brillouin-zone integration.
In the calculations, the experimental tetragonal lattice parameters were adopted, and the internal atomic coordinates within the cell were determined by the
energy minimization.

\begin{figure}[tbp]
\begin{center}
\includegraphics[angle=0,scale=0.6]{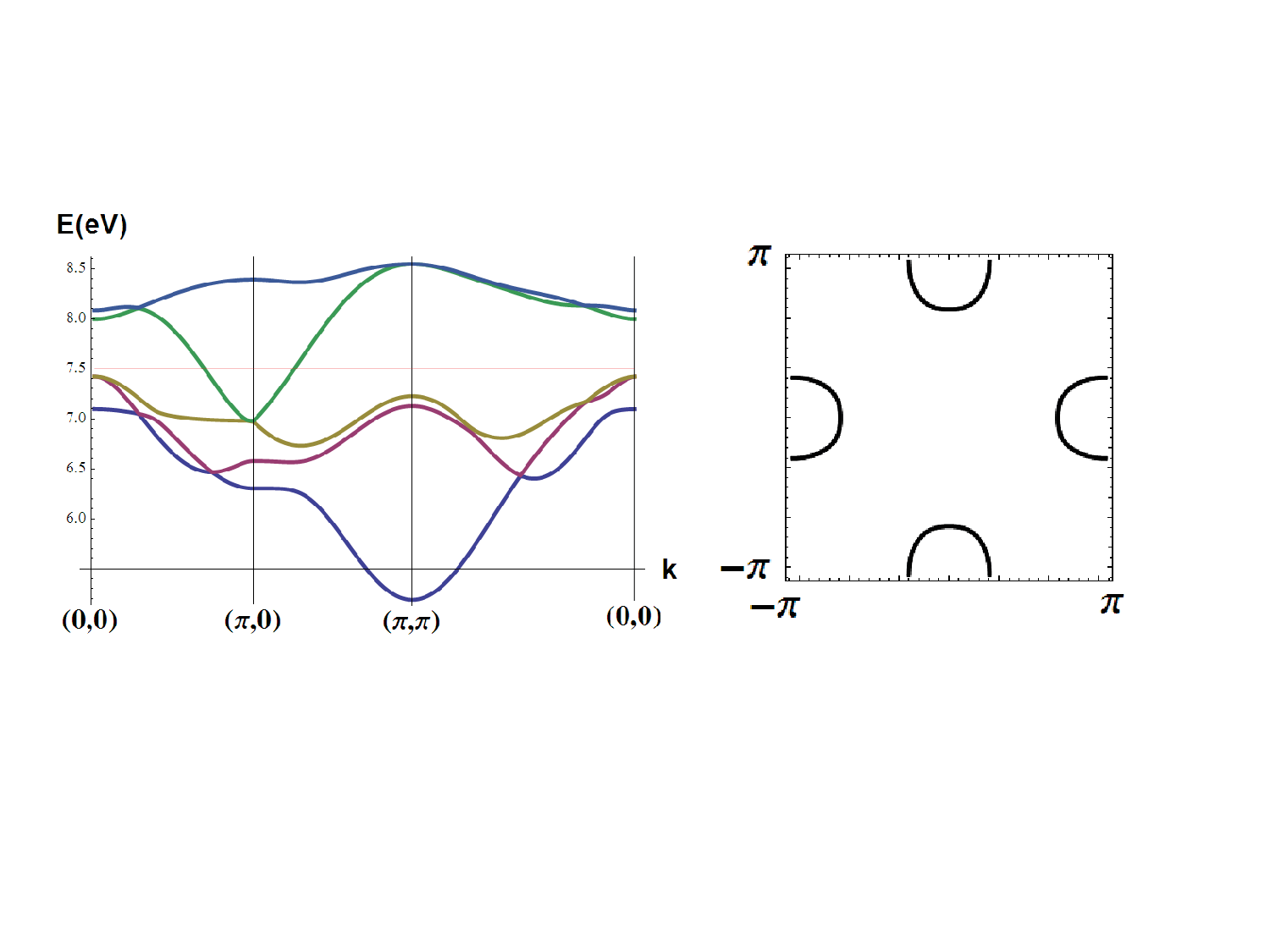}
\end{center}
\caption{(Color online) The band structure and Fermi surfaces.
Note that all the Fermi surfaces are electron like.
}
\label{band}
\end{figure}

Our bandstructure and the associated Fermi surfaces are shown in fig.~\ref{band}.
Because experimentally there is uncertainty about the Fe and K content,
we have adjusted the chemical potential so that the distance from the top of hole bands at $\Gamma$ to the Fermi energy is about $0.1$eV.
It mimics the ARPES finding of ref.~\cite{ding}.
The parameters of the tight-binding model involves the nearest- and second-neighbor hopping among the Fe 3d orbitals:
\begin{equation}
H_{\mathrm{band}}=\sum_{ij}\sum_{\alpha\beta}\sum_s
t_{ij}^{\alpha\beta} c^\dagger_{i\alpha s}c_{j\beta s}^{\vphantom{\dagger}}.
\end{equation}
Here $i,j$ labels the Fe sites, $\alpha,\beta$ label the five different Fe orbitals
($\alpha=1,..5$ denotes $Z^2$, $XZ$, $YZ$, $X^2-Y^2$, $XY$, respectively),
$s$ labels spin.
The label $X,Y$ refer to the diagonal directions,
while $x,y$ refer to the nearest neighbor Fe-Fe directions.

In unit of eV the tight-binding parameter $t_{i,i}^{\alpha\beta}$ are given by
\begin{eqnarray}
&& t_{i,i}^{\alpha\beta}=\left(
\begin{array}{ccccc}
 7.13365 & 0 & 0 & 0 & 0 \\
 0 & 7.38168 & 0 & 0 & 0 \\
 0 & 0 & 7.38168 & 0 & 0 \\
 0 & 0 & 0 & 7.31272 & 0 \\
 0 & 0 & 0 & 0 & 7.06574
\end{array}
\right)\equiv K_0^{\alpha\beta},
\\
&&t_{i,i+\hat{y}}^{\alpha\beta}=\left(
\begin{array}{ccccc}
 0.00376 & 0.07892 & 0.07892 & 0 & -0.25696 \\
 0.07892 & -0.1406 & -0.05034 & 0.18873 & 0.20995 \\
 0.07892 & -0.05034 & -0.1406 & -0.18873 & 0.20995 \\
 0 & 0.18873 & -0.18873 & -0.10725 & 0 \\
 -0.25696 & 0.20995 & 0.20995 & 0 & -0.35102
\end{array}
\right)\equiv K_1^{\alpha\beta},
\\
&&t_{i,i+\hat{x}+\hat{y}}^{\alpha\beta}=\left(
\begin{array}{ccccc}
 -0.00559 & 0 & -0.13684 & 0.11502 & 0 \\
 0 & 0.07104 & 0 & 0 & 0.05964 \\
 0.13684 & 0 & 0.23042 & 0.00961 & 0 \\
 0.11502 & 0 & -0.00961 & 0.08569 & 0 \\
 0 & -0.05964 & 0 & 0 & -0.11843
\end{array}
\right)\equiv K_2^{\alpha\beta}.
\end{eqnarray}

In terms of these parameters the $5\times 5$ Bloch Hamiltonian is given by
\begin{equation}
H(\vect{k})= M_{\mathrm{2nd}}(\vect{k})+{\frac{1}{2}}\left[M_{\mathrm{1st}}(\vect{k})R_z+R_zM_{\mathrm{1st}}^T(\vect{k})\right],
\end{equation}
where
\begin{equation}
\begin{split}
M_{\mathrm{1st}}(\vect{k})= & K_1 e^{i k_y}+e^{i k_x}R K_1R^{-1}+e^{-i k_x}R^{-1}K_1R
\\ &
+e^{-i k_y}R^{-2}K_1R^2,
\end{split}
\end{equation}
and
\begin{equation}
\begin{split}
M_{\mathrm{2nd}}(\vect{k})= & K_0+e^{i(k_x+k_y)}K_2+e^{i(k_y-k_x)}R^{-1}K_2R
\\ &
+e^{-i(k_x+k_y)}R^{-2}K_2R^2+e^{-i(k_y-k_x)}RK_2R^{-1}.
\end{split}
\end{equation}
In the above
\begin{equation}
\begin{array}{l}
R=\left(
\begin{array}{ccccc}
 1 & 0 & 0 & 0 & 0 \\
 0 & 0 & -1 & 0 & 0 \\
 0 & 1 & 0 & 0 & 0 \\
 0 & 0 & 0 & -1 & 0 \\
 0 & 0 & 0 & 0 & -1
\end{array}
\right),
\\
R_z=-\left(
\begin{array}{ccccc}
 1 & 0 & 0 & 0 & 0 \\
 0 & -1 & 0 & 0 & 0 \\
 0 & 0 & -1 & 0 & 0 \\
 0 & 0 & 0 & 1 & 0 \\
 0 & 0 & 0 & 0 & 1
\end{array}
\right),
\end{array}
\end{equation}
are the matrices representing the combined $90^\circ$ rotation and $z$ reflection and
the combined $z$ reflection and a gauge transformation on one Fe sublattice, respectively.

\begin{figure}[tbp]
\begin{center}
\includegraphics[angle=0,scale=0.6]{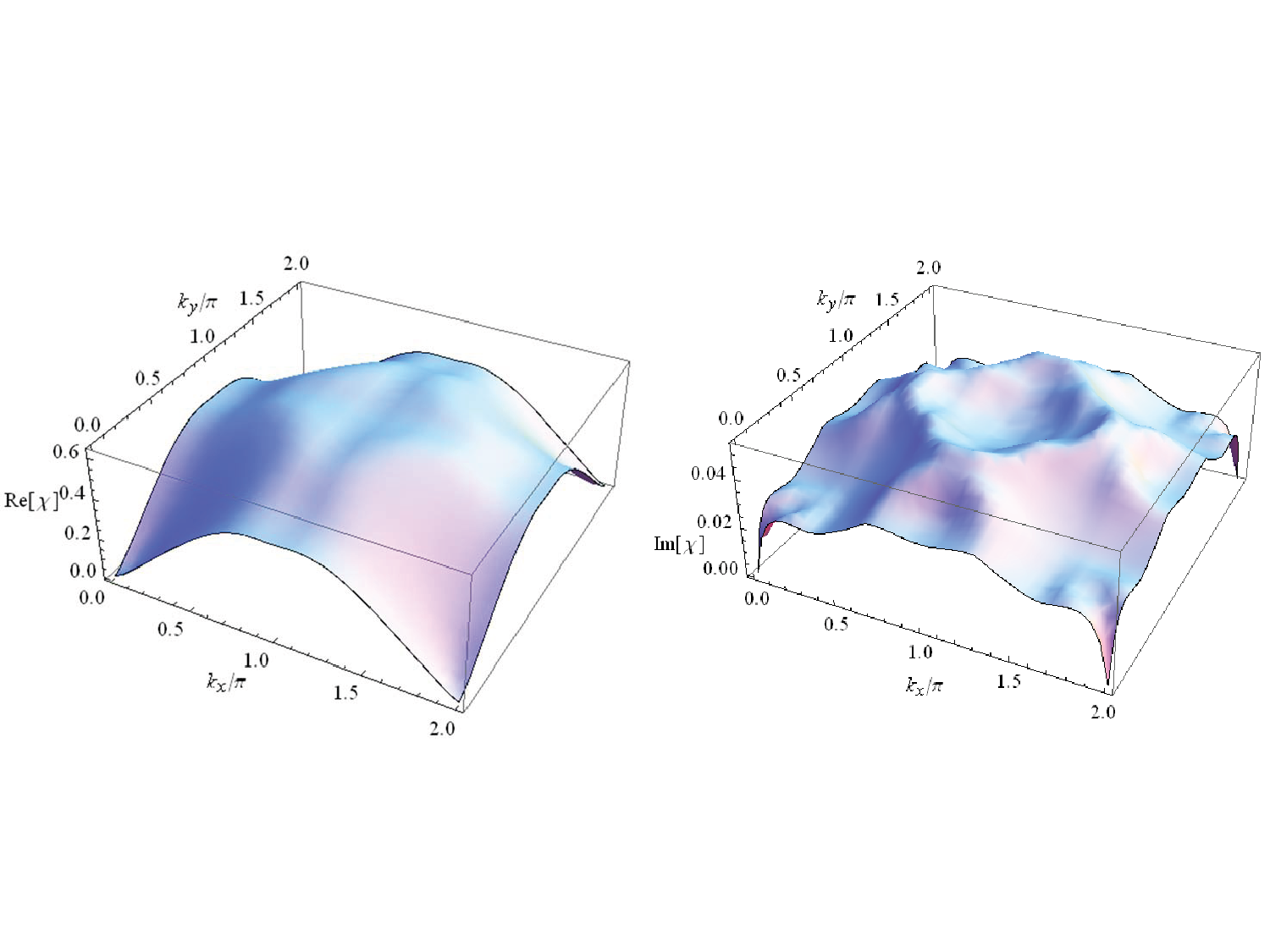}
\end{center}
\caption{(Color online)
The real(left) and imaginary(right) part of the bare spin susceptibility.
}
\label{sus}
\end{figure}

The bare magnetic susceptibility computed using the above bandstructure is shown in fig.~\ref{sus}.
A broad maximum around $\vect{Q}=(\pi,\pi)$ is found in Re($\chi$).
If we make an analogy with the pnictide superconductors, where $\vect{Q}=(\pi,0)$ or $(0,\pi)$ and the predicted pairing form factor is $s_\pm$, it is natural to guess the pairing form factor here to be $d_{x^2-y^2}$\cite{ding,thomale}.
Note that this gives rise to fully gapped superconducting state with the order parameter changing sign between the electron Fermi surfaces.
In the following we check whether this is true via a FRG calculation.

As in ref.~\cite{rg3} we model the electronic correlations by the Hubbard and Hunds types of local interactions.
The Hamiltonian is given as follows:
\begin{equation}
\begin{split}
H= &
H_{\rm band}+
U_{1}\sum_{i\alpha}n_{i\alpha\uparrow}n_{i\alpha\downarrow}+
U_{2}\sum_{i,\alpha<\beta}n_{i\alpha}n_{i\beta}
\\ &
+J_{\mathrm{H}}
\Big[
\sum_{\alpha<\beta}
\sum_{\sigma\sigma^{\prime}}
c^{+}_{i\alpha\sigma}c^{+}_{i\beta\sigma^{\prime}}
c_{i\alpha\sigma^{\prime}}c_{i\beta\sigma}
\\ &
+c^{+}_{i\alpha\uparrow}c^{+}_{i\alpha\downarrow}
c_{i\beta\downarrow}c_{i\beta\uparrow}+h.c.\Big].
\end{split}
\label{H-H-model}
\end{equation}
In the rest of the paper we use $(U_1,U_2,J_{\mathrm{H}})$=(4, 2, 0.7) eV respectively.
These value are suggested by early dynamical mean-field theory results for the pnictides\cite{DMFT}.
The sole reason for using this set of parameter is they represent intermediate couplings,
which we believe K$_x$Fe$_{2-y}$Se$_2$ is likely to be.

\section{FRG Method}
Details of the FRG method are discussed in ref.~\cite{rg3,rg8}.
We made one important modification to include the buried hole bands in the FRG calculation.
In our previous treatment bands without Fermi surface were ignored completely.
In the case of K$_x$Fe$_2$Se$_2$ (and possibly in heavily hole-doped KFe$_2$As$_2$ as well)
we believe this is no longer legitimate.
The reason is two-fold.
First consider the three different energy scales in the iron-based superconductors,
the largest energy scale is the bare bandwidth and bare Coulomb interactions of order eV,
the second energy scale is the effective spin exchange interactions about $50$meV as indicated by neutron studies\cite{spinwave},
the pairing scale(gap) is the smallest about $\sim 10$meV.
The distance between the top of the hole bands and the Fermi level in K$_x$Fe$_2$Se$_2$ is smaller than the first but comparable to the second energy scale.
Therefore for a wide range of cutoff energies virtual excitations from the hole band to above the Fermi level should contribute to the renormalization of the
effective interaction.
The second reason is that a preliminary FRG which only include the single band
that intersects the Fermi energy did not produce any strong pairing tendency,
except very weak Kohn-Luttinger-type pairing in high angular momentum channel.

Physically when the FRG cutoff energy $\Lambda$ is much larger than the distance $\Delta E_0$ between the top of the hole bands and the Fermi energy,
the fact that there are no hole pockets should not be important.
Only when
 $\Lambda\lesssim\Delta E_0$
the lack of hole pockets will play an important role.
Consequently, our FRG scheme consists of two stages.
In the first stage the FRG scheme is the same as the case where the hole pockets are present.
Thus we discretize the electron pocket band into patches represented by Fermi surfaces points as before.
For the hole bands we put fictitious ``Fermi surfaces'', small circles,
around $(0,0)$ and $(\pi,\pi)$ and discretize the Brillouin zone as if hole pockets exist.
Each (fictitious) Fermi surface is discretized into $16$ patches as in ref.~\cite{rg1,rg3,rg8}.
In the calculation of RG flow these hole band patches represented by the fictitious Fermi surface points are treated in the same way as electron band patches.
In the second stage where
 $\Lambda\lesssim\Delta E_0$
the hole bands no longer contribute virtual states in the one-loop diagrams.
When we compute pairing the fictitious Fermi surface are dropped.
Thus in the second stage of the FRG the flow of pairing channel is essentially the marginally relevant Cooper channel flow discussed by Shankar\cite{shankar}.

In our previous FRG study for systems involving both electron and hole pockets, the final effective interaction is captured reasonably well by a $J_1-J_2$ model\cite{rg8} (as far as the antiferromagnetism and pairing are concerned).
In addition, in those cases, the final effective interaction is a result of the FRG flow over the entire bandwidth, not just in slivers around the Fermi surfaces.
Thus to gain a qualitative understanding of the pairing in K$_x$Fe$_{2-y}$Se$_2$ one can use an effective $J_1-J_2$ interaction\cite{seo} to replace the FRG flow for cutoff $\Lambda >\Delta E_0$.
We have analyzed the pairing form factor using such a model and  find that for $J_1,J_2$ that are both antiferromagnetic and weak,
the nodeless $d_{x^2-y^2}$ is the dominant pairing channel for the K$_x$Fe$_2$Se$_2$ bandstructure as long as $J_1/J_2 > 0.213$.
In contrast the $s_{\pm}$ is dominant in the LaFeAsO band structure from $J_1/J_2=0$ up to $J_1/J_2\sim 1$.
Of course, in general what affects the final pairing strength and form factor involves both the strength and form of the effective interaction as well as the geometry of the Fermi surfaces.

\section{FRG Results}

\begin{figure}
\begin{center}
\includegraphics[scale=0.9]{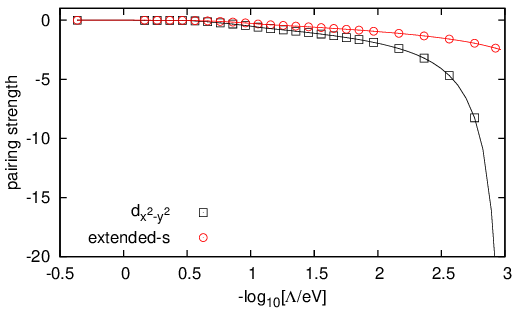}
\end{center}
\caption{(Color online)
FRG flow of two leading pairing channels for
 $(U_1,U_2,J_{\mathrm{H}})$=(4, 2, 0.7) eV.
Horizontal axis is the negative of the decadic logarithm of RG cutoff $\Lambda$.
Vertical axis is the effective pairing strength (c.f. ref.~\cite{rg8}).
}
\label{fig:flow}
\end{figure}

\begin{figure}
\begin{center}
\includegraphics[scale=0.8]{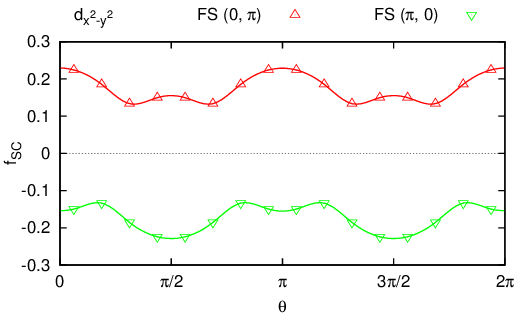}
\includegraphics[scale=0.8]{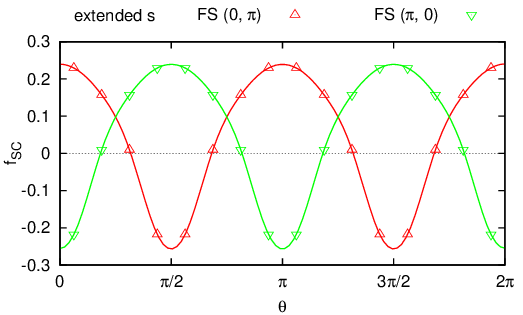}
\end{center}
\caption{(Color online)
Gap form factors (c.f. ref.~\cite{rg8}) of the two leading pairing channel for
 $(U_1,U_2,J_{\mathrm{H}})$=(4, 2, 0.7) eV.
Horizontal axis $\theta$ is the polar angle on each Fermi surface with respect to its center.
$\theta=0$ indicates the $+k_x$ direction.
Upward(red) and downward(green) triangles label the two electron Fermi surfaces at $(0,\pi)$ and $(\pi,0)$ respectively.
Top: The Leading $d_{x^2-y^2}$ gap.
Bottom: The subleading extended $s$-wave gap.
}
\label{fig:formfactor}
\end{figure}

\begin{figure}
\begin{center}
\includegraphics{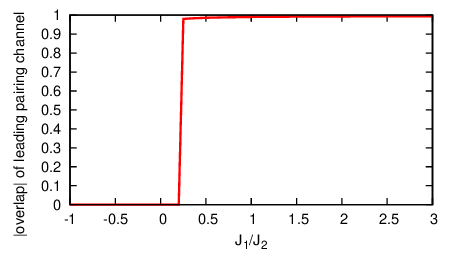}
\end{center}
\caption{(Color online)
The overlap between the gap form factors of leading pairing channels
obtained by FRG [$(U_1,U_2,J_{\mathrm{H}})$=(4, 2, 0.7) eV]
and $J_1-J_2$ mean field theory versus the ratio
$J_1/J_2$ ($J_2>0$ is antiferromagnetic).
}
\label{fig:overlap}
\end{figure}

The presentation of our results follows the format in our previous publications\cite{rg1,rg3,rg8}.
For this FRG calculation with the parameters and RG scheme described above,
we obtained divergent flow of several pairing channels,
the flow of the two leading channels are plotted in fig.~\ref{fig:flow}.
The strongest pairing channel has the  $d_{x^2-y^2}$ symmetry.
The associated divergence energy scale is $\sim 4$meV.
The second strongest pairing is a extended $s$-wave with form factor approximately described by $\cos(k_x)+\cos(k_y)$.
It has four nodes on each electron Fermi surface.
The pairing form factors of both channels are depicted in fig.~\ref{fig:formfactor}.
Interestingly both pairing channels are favored by the nearest-neighbor antiferromagnetic interactions $J_1$.
In fact as long as $J_1/J_2 > 0.2$ the $J_1-J_2$ effective interaction predicts a leading $d_{x^2-y^2}$ pairing whose form factor is almost identical to our FRG result.
The overlap between the form factor predicted by the $J_1-J_2$ model and that determined by FRG is shown in fig.~\ref{fig:overlap}.

\section{Discussions}

There are evidences from NMR that the antiferromagnetic fluctuation is weak in K$_x$Fe$_{2-y}$Se$_2$\cite{gf}.
On the surface this seems to be at odds with the notion that spin fluctuation is the main source of pairing.
However, NMR probes spin fluctuations at very low energy.
It is the high energy spin fluctuations that are good for pairing.

Obvious implications of the nodeless $d_{x^2-y^2}$ pairing is the presence of $(\pi,\pi)$ neutron resonance in the superconducting state.
In addition, phase sensitive measurement utilizing the relative orientation of crystallographic grains, or the detection of randomly trapped half flux quanta in  polycrystalline materials, can be done similar to the cuprates.
Fourier transform scanning tunneling spectroscopy should in principle reveal the enhancement of the $(\pi,\pi)$ scattering by the magnetic field.
Whether any of these is true is remained to be studied.

\begin{figure}
\begin{center}
\includegraphics[scale=0.9]{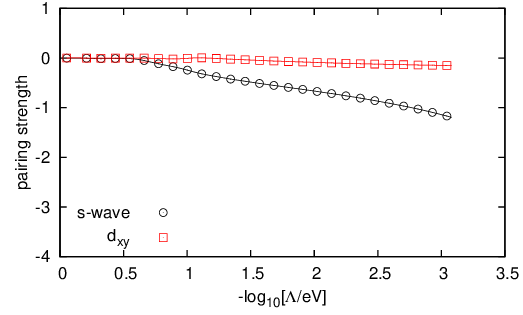}
\end{center}
\caption{(Color online)
FRG flow of two leading pairing channels for
 $(U_1,U_2,J_{\mathrm{H}})$=(4, 4, 0) eV.
Horizontal axis is the negative of the decadic logarithm of RG cutoff $\Lambda$.
Vertical axis is the effective pairing strength (c.f. ref.~\cite{rg8}).
}
\label{fig:flow440}
\end{figure}

\begin{figure}
\begin{center}
\includegraphics[scale=0.8]{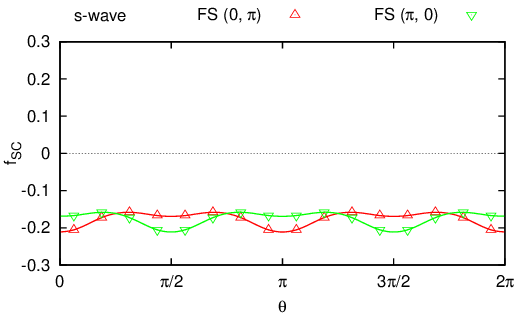}
\includegraphics[scale=0.8]{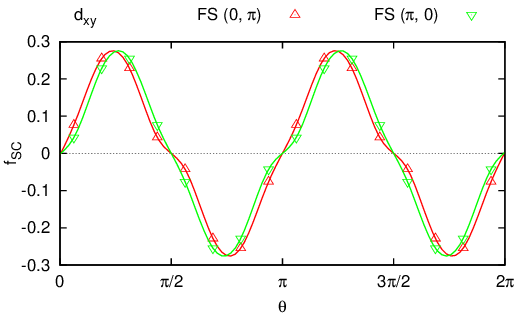}
\end{center}
\caption{(Color online)
Gap form factors (c.f. ref.~\cite{rg8}) of the two leading pairing channel for
 $(U_1,U_2,J_{\mathrm{H}})$=(4, 4, 0) eV.
Horizontal axis $\theta$ is the polar angle on each Fermi surface with respect to its center.
$\theta=0$ indicates the $+k_x$ direction.
Upward(red) and downward(green) triangles label the two electron Fermi surfaces at $(0,\pi)$ and $(\pi,0)$ respectively.
Top: The Leading $s$-wave gap.
Bottom: The subleading $d_{xy}$ gap.
}
\label{fig:formfactor440}
\end{figure}

Finally we would like to list a few caveats.
First, there is no {\it a priori} justification for ignoring the $(0,0,\pi)$ electron pocket found in the DFT calculation.
It is important to study the effect of the three dimensional electron pocket on the pairing symmetry.
Second, although the particular parameter set (including the chemical potential) we used in this study yielded $d_{x^2-y^2}$ pairing,
it is important to investigate the robustness of this pairing symmetry against chemical potential shift and/or the changing of interaction parameters.
Presently  we have studied a different set of interaction parameters $(U_1,U_2,J_{\mathrm{H}})$=(4, 4, 0) eV and found a weaker pairing divergence (fig.~\ref{fig:flow440}). More importantly the leading pairing symmetry 
becomes a fully gapped $s$-wave, and the subleading one becomes  $d_{xy}$ (fig.~\ref{fig:formfactor440}).
Both these channels may be driven by strong $J_2$.
The fully gapped $s$-wave can be viewed as the remnant of the $s_{\pm}$
pairing after the central hole pockets are removed. This result clearly calls for a more systematic study of the dependence of pairing symmetry on the parameters in our model.
Third, the real materials have Fe vacancies, and even possible partial Fe vacancy ordering.
This is obviously ignored in the current study.
Last, the effects of impurity scattering on destabilizing the $d_{x^2-y^2}$ pairing  clearly need to be addressed.

% Conclusion

In summary we have studied the pairing instabilities in K$_x$Fe$_{2-y}$Se$_2$ using a two stage functional renormalization group method.
Our calculation suggests the leading pairing channel is $d_{x^2-y^2}$ and the subleading one is extended $s$.
We find the buried hole bands make important contributions to the final effective interaction because of the closeness of their maxima to the Fermi energy.
The leading $d_{x^2-y^2}$ pairing form factor is well captured by an effective $J_1-J_2$ interaction with dominant nearest neighbor antiferromagnetic exchange.
At the end of the paper we briefly mentioned several obvious experiments that can test the $d_{x^2-y^2}$  symmetry.
We also list some caveats of the present study.

% \acknowledgments
DHL is supported by DOE grant number DE-AC02-05CH11231.
FY is supported by the NSFC Grant No.10704008.
ZYL and TX are supported by National Natural Science Foundation of China and by National Program for Basic Research of MOST, China.

\end{document}